\documentclass{article}

\usepackage{microtype}
\usepackage{graphicx}
\usepackage{subfigure}
\usepackage{booktabs} 
\usepackage{xurl}

\usepackage{hyperref}
\hypersetup{
           breaklinks=true,   
           colorlinks=true,   
           pdfusetitle=true,  
        }

\usepackage[accepted]{icml2023}

\usepackage{amsmath}
\usepackage{amssymb}
\usepackage{mathtools}
\usepackage{amsthm}

\usepackage[capitalize,noabbrev]{cleveref}

\theoremstyle{plain}

\theoremstyle{definition}

\theoremstyle{remark}

\usepackage[textsize=tiny]{todonotes}
\usepackage{csquotes}

\newcommand{\Comments}{1}
\newcommand{\mynote}[2]{\ifnum\Comments=1\textcolor{#1}{#2}\fi}
\newcommand{\mytodo}[2]{\ifnum\Comments=1%
	\todo[linecolor=#1!80!black,backgroundcolor=#1,bordercolor=#1!80!black]{#2}\fi}

\icmltitlerunning{Addressing the Risks of Bigness in Generative AI}

\begin{document}

\twocolumn[
\icmltitle{AI and the EU Digital Markets Act:\\ Addressing the Risks of Bigness in Generative AI}



\icmlsetsymbol{equal}{*}


\begin{icmlauthorlist}
\icmlauthor{Ayse Gizem Yasar}{lse}
\icmlauthor{Andrew Chong}{equal,ucb}
\icmlauthor{Evan Dong}{equal,ucornell}
\icmlauthor{Thomas Krendl Gilbert}{equal,cornellTech}
\icmlauthor{Sarah Hladikova}{equal,Tufts}
\icmlauthor{Roland Maio}{equal,cu}
\icmlauthor{Carlos Mougan}{equal,soton}
\icmlauthor{Xudong Shen}{equal,NUS}
\icmlauthor{Shubham Singh}{equal,uic}
\icmlauthor{Ana-Andreea Stoica}{equal,zzz}
\icmlauthor{Savannah Thais}{equal,cu}
\icmlauthor{Miri Zilka}{equal,Camb}

\icmlaffiliation{ucornell}{Cornell University,USA} 
\icmlaffiliation{cornellTech}{Cornell Tech, USA}
\icmlaffiliation{ucb}{University of California, Berkeley, USA}
\icmlaffiliation{cu}{Columbia University, USA}
\icmlaffiliation{lse}{ London School of Economics, London, United Kingdom}
\icmlaffiliation{soton}{University of Southampton, United Kingdom}
\icmlaffiliation{NUS}{National University of Singapore, Singapore}
\icmlaffiliation{Tufts}{Tufts University, USA}
\icmlaffiliation{uic}{University of Illinois Chicago, USA}

\icmlaffiliation{zzz}{Max Planck Institute for Intelligent Systems, T{\"u}bingen, Germany}

\icmlaffiliation{Camb}{University of Cambridge, UK}

\icmlcorrespondingauthor{Ayse Gizem Yasar}{a.g.yasar@lse.ac.uk}

\end{icmlauthorlist}

\icmlkeywords{Machine Learning, ICML}

\vskip 0.3in
]



\printAffiliationsAndNotice{\icmlEqualContribution} 
\
\begin{abstract}
As AI technology advances rapidly, concerns over the risks of bigness in digital markets are also growing. The EU's Digital Markets Act (DMA) aims to address these risks. Still, the current framework may not adequately cover generative AI systems that could become gateways for AI-based services.
This paper argues for integrating certain AI software as \enquote{core platform services} and classifying certain developers as gatekeepers under the DMA. We also propose an assessment of gatekeeper obligations to ensure they cover generative AI services. As the EU considers generative AI-specific rules and possible DMA amendments, this paper provides insights towards diversity and openness in generative AI services.~\looseness=-1
\end{abstract}

\section{Introduction}
The European Union's (EU) response to ``bigness"~\cite{brandeis1934curse,wu2018curse} in digital markets goes beyond traditional antitrust and competition law enforcement. Under EU competition law, large companies have not been viewed as inherently problematic. However, there is an increasing concern that ``a few large platforms increasingly act as gateways or gatekeepers between business users and end users and enjoy an entrenched and durable position, often as a result of the creation of conglomerate ecosystems around their core platform services, which reinforces existing entry barriers."~\cite{digitalMarketsAct}. According to the European Commission, these entrenched positions lead to unfair behaviour vis-à-vis business users of these platforms, as well as reduced innovation and contestability in core platform services. 

These concerns culminated in the creation of a new European regulation that goes beyond traditional competition law rules: the Digital Markets Act (DMA). The DMA is set up to counteract platform size and gatekeeping rather than abuse of dominance or monopoly power that competition/antitrust laws target. It aims to address the shortcomings of competition law in keeping the entry barriers low and ensuring fair game between ``gatekeepers" and their smaller rivals that depend on gatekeepers' services. The regulation covers some well-established platform services like operating systems, messaging platforms, and online advertising.

The DMA has been applauded by many who are concerned about the rise of large digital platforms. It has also been criticised because it may no longer be possible to inject diversity into digital markets once certain players have become squarely entrenched. This is a concern that we share. The DMA is silent on AI, but we observe that gatekeepers are starting to emerge in generative AI applications. The DMA provides an opportunity to maintain a fair and diverse space for AI applications before a few large players become entrenched and durable. There is now political momentum in both the EU and the US to address the emergence of generative AI gatekeepers~\cite{vestagerEU,subcommitteeAI2023}

This paper argues that the generative AI industry should be directly addressed in the DMA. In particular, we argue that generative AI services should be integrated into the DMA's list of core platform services. We show that certain generative AI services embody gatekeeper characteristics in the sense of the DMA. While certain use cases of generative AI might indirectly fall under the DMA's current list of core platform services, there are complex ways in which generative AI services may act as gatekeepers in their own right. Among those, we highlight the gatekeeping potential of generative AI service providers due to: \emph{(i)} computing power, \emph{(ii)} early mover advantages, and \emph{(iii)} data resources and integrated systems. We conclude with a discussion on how the DMA could be amended to ensure contestability and fairness in the market for generative AI services.

\section{Gatekeepers and the DMA}

The DMA came into force around the same time as the Digital Services Act, which regulates platform accountability and content moderation. Both regulations impose specific obligations on companies above certain size thresholds, albeit framed and described differently. The EU is also about to finalise its AI Act, which adopts a risk-based approach. It outright prohibits certain AI applications that bear unacceptable risks to people's safety, and introduces transparency and accountability rules for high-risk applications.

The DMA targets ``gatekeeper'' companies, which is defined under Article 3(1) as an undertaking that: \emph{(i)}~\enquote{has a significant impact on the [EU's] internal market}, \emph{(ii)}~\enquote{provides a core platform service which is an important gateway for business users to reach end-users}, and \emph{(iii)}~\enquote{enjoys an entrenched and durable position in its operations, or it is foreseeable that it will enjoy such a position in the near future.}~\looseness=-1

The current list of ``core platform services" provided in Article 2(2) covers ten established digital services, such as operating systems, web browsers, and social networking. It does not explicitly cover generative AI services. In the rest of this paper, we bring forward ways in which generative AI can be provided as a platform service and argue for its explicit integration into the DMA.

\section{Platformization of AI}
The DMA applies to core \textit{platform} services. Although some companies are likely to offer AI only as a product, another potentially effective route to profit is to provide generative AI as a platform. For example, OpenAI has created several large foundation models (e.g., GPT-4 and DALL-E) that can serve as the basis for a wide range of applications. The company began to monetize these foundational models in different ways, including: $(i)$ by releasing some models directly to the public (e.g., ChatGPT) using a \enquote{freemium} business model, and $(ii)$ by offering API access to its models and enabling the development of applications built on top of them. The latter allows organizations to integrate OpenAI's models into their own products, which they then provide to the public. To the extent generative AI applications are provided as a platform, they can be brought within the remit of the DMA.

\section{Emergence of Gatekeepers in Generative AI}

The generative AI industry is already driven by a small number of companies, notably OpenAI, Google, Microsoft, and Meta, who hold a significant competitive advantage due to their extensive data resources, specialized hardware architectures, vertical integration, network effects, financial clout, know-how, and early-mover advantage. While some of these advantages are specific to AI applications, such as specialized hardware architectures, most---such as data resources, financial clout, network effects, integration and the importance of early movers---have been present in digital markets in general and given rise to the gatekeeping positions that the DMA now seeks to address~\cite{digitalMarketsAct}. Without regulatory intervention, these significant advantages will likely turn into entrenched positions in generative AI services, as they have in other digital markets.

\textbf{Computing Power in the Hands of a Few:}
While the cost of fine-tuning large generative AI  models is decreasing, achieving state-of-the-art performance still requires a high budget, thus creating an entry barrier for potential players and inhibiting diversity and market growth. This entry barrier disproportionately affects smaller companies, public institutions, and universities, who often lack the financial resources to establish independent large-scale generative AI systems. Consequently, the concentration of cutting-edge computing power and expertise in the hands of a few players risks limiting player diversity and stifling innovation within the generative AI industry.

\textbf{Early Mover Advantage:} Early movers' head start, like OpenAI and DeepMind, in developing generative AI systems may also lead to entrenched positions.  Early movers currently face the challenge of determining whether they should make their AI systems available as open source~\cite{Vincent_2023}. However, even if they do, such open-source versions often come with strings attached to enable monetization, which was the core dispute in the European Commission's Google Android case in the context of mobile operating systems~\cite{Google_android}. In any case, monetization is now becoming increasingly common for developers ~\cite{reutersChatGPT}. 

\textbf{Data Resources and Integrated Systems: }
A new trend in generative AI systems is the recent influx of integrated services: search engines integrate LLMs, personal assistants, note-taking, editing, creative tasks automation, video-editing applications, or generative AI-augmented search~\cite{googleSupercharging}. 
As integrated systems that use generative AI become ubiquitous, their convenience and potential for creative endeavours trade off with user autonomy. Meanwhile, platform tendency for integration and the associated loss of user autonomy is not new: Google, Microsoft and Apple have all pushed for integrated systems in their now-established services. 
Furthermore, large players may benefit from both existing troves of data from legacy applications (e.g., Google from G-suite users), as well as data generated as users interact with AI applications (through prompts and other inputs). The more popular an application is, the more it will benefit from human feedback, which creates feedback loops associated with network effects---another factor motivating the DMA. As a result of service and data integration, it becomes increasingly difficult for a user to switch between platforms and transport their data and projects. The DMA specifically prohibits data and service integration across different offerings of core platform services (e.g., Article 5(2)(b) and 5(8)). However, as mentioned above, this list does not include any generative AI applications.


\section{Contestability and Fairness in the Generative AI Industry}

The potential harms of AI have been discussed for years, but the current race toward generative AI enhancement might lead to the emergence of a few large players at the expense of contestability and fairness in the generative AI industry. The draft AI Act and the DMA do not directly address this problem. 


Certain cases of generative AI applications might indirectly fall within the remit of the DMA when gatekeepers integrate proprietary or third-party generative AI into their core platform services. However, such indirect application falls short of addressing our concerns. The DMA’s gatekeeper obligations that are designed to counteract contestability and fairness issues will not apply to their generative AI offering when it is provided as a standalone platform service.

There is evidence that unfair practices are already emerging in generative AI. For example, Microsoft has reportedly threatened to cut off the access of at least two of its search index business customers that were building their own generative AI tools using data from the search index~\cite{microsoftdata}. Such behavior appears to be intended to weaken the contestability of Microsoft’s own generative AI offering by denying potential competitors the key input factor of data. Such practices are exactly what the DMA aims to eliminate, but they are currently not caught by the regulation. Generative AI providers remain free to leverage their superior bargaining power to engage in such unfair practices, and their already considerable economic power to prevent contestability. 

\looseness=-1
The DMA provides an opportunity to address contestability and fairness issues in the generative AI industry by designating certain AI software as core platform services and certain developers as gatekeepers. An initial assessment of gatekeeper obligations under the DMA (Articles 5-6-7) reveals that some of them would already apply to generative AI systems. For example, under Article 6(2), gatekeepers are prevented from using, in competition with their business users, the data generated by business users of their core platform services and by these businesses’ customers. This provision would prevent generative AI gatekeepers from free-riding on data produced by businesses relying on their API to provide downstream applications, either in generative AI verticals or other industries. Similarly, a generative AI gatekeeper would be prevented from forcing business users to rely on its own identification services, web browser engine, payment service, or other technical services, for services provided using the gatekeeper’s generative AI under Article 5(7). A similar provision under Article 5(8) would prevent generative AI gatekeepers from forcing businesses and end users to subscribe to or register with any of its other core platform services as a condition to use its generative AI service.

The DMA can thus support contestability and innovation in AI systems, promoting the development of a more diverse and accessible generative AI market.

In conclusion, this paper presents a set of proposals to amend the DMA to prevent the emergence of entrenched positions and to address potential harms related to gatekeeping in the generative AI industry.

\newpage
\onecolumn
\bibliography{bib}

\begin{thebibliography}{10}
\providecommand{\natexlab}[1]{#1}
\providecommand{\url}[1]{\texttt{#1}}
\expandafter\ifx\csname urlstyle\endcsname\relax
  \providecommand{\doi}[1]{doi: #1}\else
  \providecommand{\doi}{doi: \begingroup \urlstyle{rm}\Url}\fi

\bibitem[Brandeis(1934)]{brandeis1934curse}
Brandeis, L.~D.
\newblock \emph{The Curse of Bigness: Miscellaneous Papers of Louis D.
  Brandeis}.
\newblock Viking Press, 1934.

\bibitem[Dastin et~al.(2023)Dastin, Hu, and Dave]{reutersChatGPT}
Dastin, J., Hu, K., and Dave, P.
\newblock Exclusive: Chatgpt owner openai projects \$1 billion in revenue by
  2024.
\newblock Reuters, 2023.
\newblock URL
  \url{https://www.reuters.com/business/chatgpt-owner-openai-projects-1-billion-revenue-by-2024-sources-2022-12-15/}.

\bibitem[{European Commission}(2020)]{digitalMarketsAct}
{European Commission}.
\newblock Proposal for a regulation of the european parliament and of the
  council on contestable and fair markets in the digital sector (digital
  markets act).
\newblock COM/2020/842 final, 2020.
\newblock URL
  \url{https://eur-lex.europa.eu/legal-content/en/TXT/?uri=COM%3A2020%3A842%3AFIN}.

\bibitem[{European Commission}(2022)]{Google_android}
{European Commission}.
\newblock Google and {Alphabet} v {Commission} ({Google} {Android}), 2022.
\newblock URL
  \url{https://ec.europa.eu/competition/antitrust/cases/dec_docs/40099/40099_9993_3.pdf}.

\bibitem[Nylen \& Bass(2023)Nylen and Bass]{microsoftdata}
Nylen, L. and Bass, D.
\newblock Microsoft threatens data restrictions in rival ai search.
\newblock Bloomberg, 2023.
\newblock URL
  \url{https://finance.yahoo.com/news/microsoft-threatens-restrict-data-rival-002746878.html}.

\bibitem[Reid(2023)]{googleSupercharging}
Reid, E.
\newblock Supercharging search with generative ai.
\newblock Google, 2023.
\newblock URL \url{https://blog.google/products/search/generative-ai-search/}.

\bibitem[{Subcommittee on Privacy, Technology, and the
  Law}(2023)]{subcommitteeAI2023}
{Subcommittee on Privacy, Technology, and the Law}.
\newblock Oversight of a.i.: Rules for artificial intelligence.
\newblock Subcommittee Hearing, 2023.
\newblock URL
  \url{https://www.judiciary.senate.gov/committee-activity/hearings/oversight-of-ai-rules-for-artificial-intelligence}.
\newblock Dirksen Senate Office Building Room 226, Washington, D.C.

\bibitem[Vestager(2023)]{vestagerEU}
Vestager, M.
\newblock Eu eyes new rules for generative ai this year.
\newblock Nikkei Asie Interview, 2023.
\newblock URL
  \url{https://asia.nikkei.com/Editor-s-Picks/Interview/EU-eyes-new-rules-for-generative-AI-this-year-Vestager}.

\bibitem[Vincent(2023)]{Vincent_2023}
Vincent, J.
\newblock Openai co-founder on company’s past approach to openly sharing
  research: “we were wrong”, Mar 2023.
\newblock URL
  \url{https://www.theverge.com/2023/3/15/23640180/openai-gpt-4-launch-closed-research-ilya-sutskever-interview}.

\bibitem[Wu(2018)]{wu2018curse}
Wu, T.
\newblock The curse of bigness.
\newblock \emph{Columbia Global Reports}, 75, 2018.

\end{thebibliography}
\bibliographystyle{icml2023}

\newpage
\appendix
\onecolumn


\end{document}